\documentclass[useAMS,usenatbib]{mn2e}

\usepackage{graphicx}
\usepackage{amssymb}
\usepackage{amsmath}

\newcommand{\apj}{ApJ}
\newcommand{\apjs}{ApJS}
\newcommand{\apjl}{ApJ}
\newcommand{\mnras}{MNRAS}
\newcommand{\aap}{A\&A}

\newcommand{\pasp}{PASP}

\begin{document}

\title[Effects of Field-Aligned Rotation]
   {Dynamical Simulations of Magnetically Channeled Line-Driven Stellar Winds:
\\  II. The Effects of Field-Aligned Rotation
}
 \author[A. ud-Doula et al.]
 {Asif ud-Doula\thanks{Also at the Department of Physics and Astronomy,
Swarthmore College,
Swarthmore, PA 19081}
 , Stanley P. Owocki and Richard H.D. Townsend \\
 Bartol Research Institute,
     University of Delaware,
     Newark, DE 19716
}

\date{\today}

\def\<<{{\ll}}
\def\>>{{\gg}}
\def\wig{{\sim}}
\def\spose#1{\hbox to 0pt{#1\hss}}
\def\ltwig{\mathrel{\spose{\lower 3pt\hbox{$\mathchar"218$}}
     \raise 2.0pt\hbox{$\mathchar"13C$}}}
\def\gtwig{\mathrel{\spose{\lower 3pt\hbox{$\mathchar"218$}}
     \raise 2.0pt\hbox{$\mathchar"13E$}}}
\def\+/-{{\pm}}
\def\=={{\equiv}}
\def\Rstar{R_{\ast}}
\def\Mstar{M_{\ast}}
\def\Lstar{L_{\ast}}
\def\Tstar{T_{\ast}}
\def\gstar{g_{\ast}}
\def\vth{v_{th}}
\def\grad{g_{rad}}
\def\glines{g_{\rm{lines}}}
\def\Mdot{\dot M}
\def\mdot{\dot m}
\def\yr{{\rm yr}}
\def\ksec{{\rm ksec}}
\def\kms{{\rm km/s}}
\def\qad{\dot q_{ad}}
\def\qlines{\dot q_{lines}}
\def\solar{\odot}
\def\Msun{M_{\solar}}
\def\msbyr{\Msun/\yr}
\def\Rsun{R_{\solar}}
\def\Lsun{L_{\solar}}
\def\Be{{\rm Be}}
\def\Rpole{R_{p}}
\def\Req{R_{eq}}
\def\Rmin{R_{min}}
\def\Rmax{R_{max}}
\def\Rstag{R_{stag}}
\def\vinf{V_\infty}
\def\Vrot{V_{\rm{rot}}}
\def\Vcrit{V_{crit}}
\def\half{{1 \over 2}}
\newcommand{\beq}{\begin{equation}}
\newcommand{\eeq}{\end{equation}}
\newcommand{\beqa}{\begin{eqnarray}}
\newcommand{\eeqa}{\end{eqnarray}}
\def\phip{{\phi'}}

\maketitle

\begin{abstract}
Building upon our previous MHD simulation study of magnetic channeling
in radiatively driven stellar winds, we examine here the additional
dynamical effects of stellar {\em rotation} in the (still) 2-D axisymmetric
case of an aligned dipole surface field.
In addition to the magnetic confinement parameter $\eta_{\ast}$
introduced in Paper I, we characterize the stellar rotation in terms of a
parameter  $W \equiv V_{\rm{rot}}/V_{\rm{orb}}$ (the ratio of the equatorial surface
rotation speed to orbital speed), examining specifically models with moderately
strong rotation $W =$~0.25 and 0.5, and comparing these to analogous
non-rotating cases.
Defining the associated Alfv\'{e}n radius
$R_{\rm{A}} \approx \eta_{\ast}^{1/4} \Rstar$
and Kepler corotation radius $R_{\rm{K}} \approx W^{-2/3} \Rstar$,
we find rotation effects are weak for models with $R_{\rm{A}} < R_{\rm{K}}$,
but can be substantial and even dominant for models with $R_{\rm{A}} \gtwig
R_{\rm{K}}$.
In particular, by extending our simulations to magnetic confinement
parameters (up to $\eta_{\ast} = 1000$)
that are well above those ($\eta_{\ast} = 10$)
considered in Paper I, we are able to study cases with $R_{\rm{A}} \gg
R_{\rm{K}}$;
we find that these do indeed show clear formation of the
{\em rigid-body} disk predicted in previous analytic models,
with however a rather complex, dynamic behavior characterized by both
episodes of downward infall and outward breakout that limit the
buildup of disk mass.
Overall, the results provide an intriguing glimpse into the complex
interplay between rotation and magnetic confinement, and form the
basis for a full MHD description of the rigid-body disks expected in
strongly magnetic Bp stars like $\sigma$~Ori~E.

\end{abstract}

\begin{keywords}
MHD ---
Stars: winds ---
Stars: magnetic fields ---
Stars: early-type ---
Stars: rotation ---
Stars: mass loss
\end{keywords}

\section{INTRODUCTION}

In Paper I of this series \citep{udDOwo2002}, we examined the
effect of a large-scale dipole magnetic field on the radiatively driven wind
from a hot, massive star.
The radiative envelopes of such hot stars means they lack the
strong convection zone that drives the dynamo generation of magnetic
activity cycles in the Sun and other relatively cool stars.
Nonetheless, in recent years spectro-polarimetric observations have
led to positive detections of large-scale fields in
several such hot stars, often well fit by a dipole tilted relative
to the star's rotation axis
(e.g., \citealt{Don2002}).
In some cases the associated period of rotational modulation
is quite long, weeks or even years
(e.g. in $\theta^{1}$~Ori~C,  HD191612;  \citealt{Don2006}),
implying that the direct {\em dynamical} effect of rotation
on the magnetic channeling of the wind
is likely to be limited.
As a first approximation, the MHD simulation models of Paper I thus
ignored the effects of rotation.

More generally, however, massive stars tend to have quite rapid rotation,
as evidenced both by the substantial broadening in photospheric spectral
lines \citep{ConEbb1977,Fuk1982}, which indicate projected
rotation speeds of hundreds of km/s, and by the
relatively short-period of observed modulations for some stars,
e.g. the magnetic Bp star $\sigma$~Ori~E, for which the inferred
rotation period is about 1.2~d
\citep{Wal1981}.
Both lines of evidence suggest that hot-star rotation rates are commonly a
substantial fraction of the `critical' rate
at which the equatorial surface would be in Keplerian orbit.
Since this implies centrifugal forces that are comparable to the inward pull
of gravity, it is clear that such levels of rotation could
significantly influence the  magnetic channeling of a stellar wind.

Previous studies have focused in particular on the potential role of
magnetic fields in spinning up the wind outflow and channeling it into
an equatorial disk that might be centrifugally supported against gravity.
\citet{Cas2002} argued that such magnetic spin-up could effectively eject material into
a ``Magnetically Torqued Disk'' (MTD), in which individual fluid elements
would be in local Keplerian orbit, and
specifically proposed this as a model for the Keplerian ``decretion disks''
inferred for Be stars.
For the chemically peculiar Bp stars that have been directly observed
to have very strong magnetic fields ($\gtwig 10^{4}$~G),
\citet[herefter TO-05]{TowOwo2005}  developed a somewhat
different ``Rigidly Rotating Magnetosphere'' (RRM) paradigm
in which the field  again spins up and channels wind material into a
disk, but now is also sufficiently strong to hold it in
{\em rigid body} rotation.
This RRM model has proven particularly successful in explaining the
rotationally modulated Balmer emission observed from
$\sigma$~Ori~E
\citep*{Tow2005}.

To test these semi-analytic paradigms, we have made some initial
efforts  to extend the numerical MHD simulations of Paper I to include
rotation in the simple 2D axisymmetric case of a rotation-aligned
dipole.
The results indicate that a large-scale field strong enough to torque
wind material to Keplerian orbital speed tends also to propel material away
from the star, rather than into the kind of stationary, Keplerian disk
envisioned in the MTD model
\citep*{udD2006,Owo2006a,Owo2006b}.
However, for cases with  strong-enough magnetic confinement to hold
material down against such outward escape, there can indeed form a
limited rigid-body disk quite similar to that predicted by the RRM model
\citep{OwoudD2003,Owo2006}.
But even in such cases, there is irregular breakout of material from the outer
disk, leading to a sudden magnetic reconnection heating that could explain
the hard X-ray flares seen from $\sigma$~Ori~E
\citep{udD2006}.

The present paper further extends these previous MHD simulations with
a more extensive 2-D parameter study covering models over a range
in both rotation rate and degree of magnetic confinement.
A particular focus is to develop a clearer physical picture of the
complex competition between wind-fed build-up of material in a disk vs. losses
by both outward ejection and infall back to the star.\footnote{ To
allow focus on these issues of disk buildup, we  defer here any discussion
of wind angular momentum loss and the resulting stellar spindown to a
future, follow-up paper.}
Moreover, the broad parameter study here allows us to examine in
detail how these processes are affected by various combinations of
rotation rate and magnetic field strength.
To lay the basis for the results presented in section 4,
section 2 first reviews the general numerical MHD approach,
and section 3 defines the overall parameter domain.
Section 5 concludes with a summary and outline for future work.
\section{NUMERICAL METHOD}

\subsection{Vector form of basic MHD equations}

As in Paper I, our general approach is to use the ZEUS-3D
\citep{StoNor1992} numerical MHD code  to evolve a consistent dynamical
solution for a line-driven stellar wind
from a star with a dipole surface field.
Our implementation here again adopts spherical polar coordinates with
radius $r$, co-latitude $\theta$, and azimuth $\phi$, but now in a
``2.5-D'' formulation that allows for non-zero azimuthal components of
both the magnetic field $B_{\phi}$ and velocity $v_{\phi}$, while still
assuming all quantities are constant in the azimuthal coordinate angle $\phi$.
To maintain this 2.5-D axisymmetry, we assume the stellar magnetic field
to be a pure-dipole with polar axis aligned with the rotation axis of
the star.

In vector form, the standard formulation of magnetohydrodynamics
includes equations for mass continuity,
\begin{equation}
 \frac{\partial \rho}{\partial t} +
 {\bf v} \cdot \nabla \rho
 + \rho \nabla \cdot {\bf v} = 0 \, ,
\label{masscon}
\end{equation}
and momentum balance,
\begin{equation}
\frac{\partial {\bf v}}{\partial t} +
{\bf v} \cdot \nabla {\bf v}
= - \frac{\nabla p}{\rho} +\frac{1}{4 \pi \rho} (\nabla \times
{\bf B}) \times {\bf B} - { GM
{\hat {\bf r}} \over r^2 }
+ {\bf g}_{\rm{lines}}
,
\label{eom}
\end{equation}
where the notation follows common conventions,
and is defined in detail in section 2 of Paper I.
(Note that eqn. (\ref{eom}) here corrects some minor errors in the
corresponding eqn. (2) of Paper I.)

\subsection{Rotation terms in the advective acceleration}

The inclusion of a finite rotation in the present work
leads to additional nonzero
terms proportional to the azimuthal velocity $v_{\phi}$ within the
three vector components of the advective acceleration,
\beq
\label {advderivx}
\left [ {\bf v} \cdot \nabla {\bf v} \right ]_{r} =
{ v_r } \, { \partial v_r \over \partial r  } ~ + ~
{ v_\theta \over r} \, { \partial v_r \over \partial \theta  }
-
{ v_\theta^2  + v_\phi^2 \over r }
\, ,
\eeq
\beq
\label {advderivy}
\left [ {\bf v} \cdot \nabla {\bf v} \right ]_{\theta} =
{ v_r } \, {\partial v_\theta \over \partial r } ~ + ~
{ v_\theta \over r } \, { \partial v_\theta \over \partial \theta  }
-
\cot \theta \, {v_\phi^2 \over r} ~ + ~
{ v_r v_\theta \over r }
\, ,
\eeq
\beq
\label {advderivz}
\left [ {\bf v} \cdot \nabla {\bf v} \right ]_{\phi} =
{ v_r } \, {\partial v_\phi \over \partial r } ~ + ~
{ v_\theta \over r } \, { \partial v_\phi \over \partial \theta  }
~ + ~ \cot \theta \, {v_\phi v_\theta \over r}
~ + ~ { v_\phi v_r \over r } \,
 .
\eeq
The terms here without derivatives represent the inertial forces
arising from the curvature of the coordinate system, namely centrifugal and coriolis forces,
which, e.g., in the absence of external torques, enforce conservation of
angular momentum within the rotating flow  described in spherical
coordinates.

Of course, in magnetic models the Lorentz force, represented by the
second term on the right side of eqn. (\ref{eom}), can significantly
channel the flow, competing against these inertial terms.
In rotating models, this Lorentz force can now also impart a
significant torque to {\em spin up} the outflow;
moreover it now includes additional terms proportional to the
non-zero $B_{\phi}$, which themselves represent a component for outward
angular momentum transport.

This competition between the magnetic Lorentz force and the inertia
terms associated with {\em rotation} represents a central focus of the
present study.
In some ways, it parallels the central competition examined in Paper
1, namely between magnetic forces and the inertia associated with
the {\em radial} outflow of the wind.

\subsection{Radial driving of wind outflow}

This radial outflow arises from the strong radial driving of the line-force,
$\bf{g}_{\rm{lines}}$.
As in Paper I, we model this here in terms of the
standard  \citet*[hereafter CAK]{Cas1975} formalism,
corrected for the finite cone angle of the star, using
a spherical expansion approximation for the local flow gradients
\citep*{Pau1985,FriAbb1986}
and ignoring {\em non-radial} line-force components that can
arise in a non-spherical outflow.
Although such
non-radial
terms are typically only a few percent of the
radial force, in non-magnetic models of rotating winds, they act without much
competition in the lateral force balance, and so can have surprisingly
strong effects on the wind channeling and rotation
\citep*{OwoCra1996,GayOwo2000}.
But in magnetic models with an already strong component of non-radial
force, such terms are not very significant, and
since their full inclusion substantially complicates both the
numerical computation and the analysis of simulation results, we have elected to
defer further consideration of such non-radial line-force terms to future studies.

By limiting our study to moderately fast rotation, half
or less of the critical rate, we are also able to neglect the effects
of stellar oblateness and gravity darkening.

\subsection{Isothermal flow approximation}

Another simplification retained from Paper I is that the flow is
strictly isothermal.
This is generally a reasonable approximation in steady-state, spherical
wind models, wherein the competition between photoionization heating
and radiative cooling keeps the wind close to the stellar effective
temperature \citep{Pau1987,Dre1989}.
However, models with significant magnetic channeling can guide the flow toward
strong shock compressions that heat the gas to temperatures of
millions of Kelvin. The associated extensive X-ray emission
\citep{BabMon1997a,BabMon1997b}
has indeed been a focus of some our
previous simulations aimed at modeling the observed X-ray spectrum from
$\theta^{1}$~Ori C \citep{Gag2005}.
Moreover, our other simulations show that reconnection heating associated
with centrifugally driven breakout events might provide a basis for
explaining the relatively hard X-ray flare events seen in the magnetic
B-star $\sigma$~Ori~E \citep{udD2006}.

While such detailed treatments of the wind energy balance can thus be
quite important for modeling the X-ray emission from specific stars,
including this in the general parameter study here would require introducing
an additional free parameter, associated with the wind density, and
representing the relative importance of radiative cooling in the
post-shock region \citep{udD2003}.
This would in effect require a {\em three}-dimensional parameter study,
representing cooling, magnetic confinement, and rotation.
To maintain the focus here on just the one additional degree of freedom
associated with rotation, built upon the study of isothermal
magnetic confinement in Paper I, we again assume a simple isothermal wind
with the wind temperature kept equal to the stellar effective temperature,
here taken to be 50,000~K.

A further advantage is that, at such a temperature, the gas
pressure terms in eqn. (\ref{eom}) are typically unimportant throughout
most of the supersonic outflow.
Thus, although these pressure terms are still fully included in the
numerical simulations, they can be largely ignored in interpretation
of results, allowing for a focus on the dominant competing forces
associated with gravity, radiative driving, magnetic field and flow
inertia.

\subsection{Boundary conditions and numerical considerations}

Finally, numerical specifications such as the computational grid and
boundary conditions are again similar to Paper I, except of course
that the lower boundary now has a non-zero azimuthal speed
$v_{\phi} (\Rstar, \theta) = V_{\rm{rot}} \sin \theta$,
where $V_{\rm{rot}}$ is the equatorial surface rotation speed.
Also, instead of setting the azimuthal field to zero at the stellar
surface, as in Paper I, we now compute a generally non-zero $B_{\phi}$
at the lower boundary ghost-zone by linear extrapolation from the values in
the two innermost zones of the actual computational grid.
This assumes vanishing second derivatives of the azimuthal field
components.

{
In all simulations presented here the time step is based on relatively low
Courant number of 0.3, a choice that helps ensure stability and
reduced error in computed shock properties \citep{Falle2002}.
Simulations of selected models with an even lower Courant number of 0.1
gave very similar results to the standard runs.
}

\section{TWO-PARAMETER STUDY}

\subsection{Magnetic confinement parameter $\eta_{\ast}$}
\label{mcpsec}

Let us now consider how best to frame our parameter study for
the combined effects of
rotation and magnetic channeling in a line-driven wind.
In the absence of significant  rotation or magnetic fields,
the line-force overcomes the stellar gravity to drive a nearly
radial wind outflow characterized by a
mass loss rate ${\dot M}$  and terminal wind speed $\vinf$.
When a magnetic field is added, the inertia of this radial outflow
competes against the Lorentz forces.
A key result of Paper I is that the overall net effect of a magnetic
field in diverting such a wind outflow can be characterized by a single
{\em magnetic confinement parameter},
\beq
\eta_{\ast} \equiv \frac {B_{eq}^2 \, \Rstar^2} {\dot{M} \, \vinf}
\, ,
\label{esdef}
\eeq
where $B_{eq}$ is the surface field strength at the magnetic equator.
This sets the
scale of the ratio of magnetic energy to wind kinetic energy,
\beq
\eta(r)\equiv \frac {B^2/8\pi}{\rho v^2/2}
=
\eta_{\ast} \,
\left [\frac { (r/\Rstar)^{2-2q}}{(1- \Rstar/r)^{\beta}} \right ]
= \left ( \frac{V_{A}}{v} \right )^{2}
= M_{A}^{-2}
\, .
\label{eta-eqn}
\eeq
The last two equalities emphasize this energy ratio can also be cast as the
square of the ratio of the Alfv\'{e}n speed, $V_{A} \equiv B/\sqrt{4
\pi \rho}$, to flow speed, $v$,  i.e. as the inverse square of the
Alfv\'{e}nic Mach number, $M_{A} \equiv v/V_{A}$.

The square bracket factor in the middle equality shows the overall
radial variation;
$q$ is the power-law exponent for radial decline of
the assumed stellar field, e.g. $q=3$ for a pure dipole,
and $\beta$ is the velocity-law index, with typically $\beta \approx 1$.
For a star with a non-zero field, we have $\eta_{\ast} > 0$, and so given
the vanishing of the flow speed at the atmospheric wind base, this
energy ratio always starts as a large number near the stellar surface,
$\eta(r \rightarrow \Rstar) \rightarrow \infty$.
But from there outward it declines quite steeply, asymptotically as
$r^{-4}$ for a dipole, crossing unity at the
Alfv\'{e}n radius defined implicitly by  $\eta(R_A) \equiv 1$.

For a canonical $\beta=1$ wind velocity law,
explicit  solution for $R_{\rm{A}}$ along the magnetic equator
requires finding the appropriate root of
\begin{equation}
    \left ( {R_A \over R_\ast } \right )^{2q-2} -
    \left ( {R_A \over R_\ast } \right )^{2q-3} = \eta_{\ast}
\, ,
\label{radef}
\end{equation}
which for integer $2q$ is just a simple polynomial, specifically
a quadratic, cubic, or quartic for $q =$~2, 2.5, or 3.
Even for non-integer values of $2q$, the relevant solutions
can be approximated (via numerical fitting)
to within a few percent by the simple general expression,
\beq
\frac{R_{\rm{A}}}{\Rstar}
\approx 1  + (\eta_{\ast} + 1/4)^{1/(2q-2)} -  (1/4)^{1/(2q-2)}
\, .
\label{raapp}
\eeq
For weak confinement, $\eta_{\ast} \ll 1$, we find
$R_{\rm{A}} \rightarrow \Rstar$,
while for strong confinement, $\eta_{\ast} \gg 1$, we obtain
$R_{\rm{A}} \rightarrow \eta_{\ast}^{1/(2q-2)} \Rstar$.
In particular, for the standard dipole case with $q=3$,
we expect the strong-confinement scaling
$R_{\rm{A}}/R_{\ast} \approx 0.3+\eta_{\ast}^{1/4}$.

Clearly $R_{\rm{A}}$ represents the radius at which the wind speed $v$
exceeds the local Alfv\'{e}n speed $V_{A}$.
But Paper I showed that it also characterizes the maximum radius where the
magnetic field still dominates over the wind, and is just somewhat
above (i.e., by 20-30\%) the maximum extent
of closed loops in the magnetosphere.
Moreover, as we shall see below, in rotating winds these closed loop regions
tend to co-rotate nearly rigidly with the underlying star, and in this
sense $R_{\rm{A}}$ is just above the maximum radius for wind co-rotation
near the equator.
For convenience in discussing results, let us thus denote the maximum
radius of such closed (and generally co-rotating) loops as
\beq
R_{\rm{c}} \approx \Rstar +  0.7 ( R_{\rm{A}} - \Rstar )
\, .
\label{rcdef}
\eeq

\subsection{Rotation parameter $W$}
\label{wdefsec}

Let us next seek a similarly convenient parameterization for the stellar
rotation.
This can again be characterized in terms of a speed, namely the
equatorial surface rotation speed $V_{\rm{rot}}$.
But instead of relating that to the flow speed or Alfv\'{e}n speed
in the stellar {\em wind},
the stellar origin of rotation suggests it may be better to compare
it to a speed representative of the gravity at the stellar {\em surface}.
Specifically, let us thus define our dimensionless rotation parameter as
\beq
W \equiv \frac{ V_{\rm{rot}}} { V_{\rm{orb}}}\, ,
\label{wdef}
\eeq
where $V_{\rm{orb}} \equiv \sqrt {GM /\Rstar}$
is the {\em orbital} speed near the equatorial surface\footnote{
This is closely related to the commonly used rotation parameter
$\omega \equiv \Omega/\Omega_{crit}$, defined by the
star's angular rotating frequency $\Omega$ relative to the value this
would have as the star approaches ``critical'' rotation, $\Omega_{c}$.
Our choice here more directly relates to the additional local speed needed to
propel material into Keplerian orbit, and avoids some subtle assumptions
(e.g. rigid-body rotation using a Roche potential for gravity)
about how the global stellar envelope structure adjusts to approaching
the critical rotation limit.
}.
This characterizes the azimuthal speed needed for the
outward centrifugal forces to balance the stellar surface gravity.
It is only a factor $1/\sqrt{2}$ less than the speed $V_{\rm{esc}}$
needed to fully {\em escape} the star's surface gravitational potential.

For a non-magnetic rotating star, conservation of angular momentum in
a wind outflow causes the azimuthal speed near the equator to decline
outward as $v_{\phi} \sim 1/r$, meaning that rotation effects tend to be of
diminishing importance in the outer wind.

By contrast, in a rotating star with a sufficiently strong magnetic field,
magnetic torques on the wind can spin it up;
for some region near the star, i.e.,
up to about the maximum loop closure radius $R_{\rm{c}}$,
they can even maintain a nearly rigid-body rotation,
for which the azimuthal speed now
{\em increases} outward in proportion to the radius,
\beq
v_{\phi} (r) = V_{\rm{rot}} \, \frac{r}{\Rstar} ~~ ; ~~ r \ltwig R_{\rm{c}}
\, .
\label{vprig}
\eeq
As such, even for a star with surface rotation below the orbital
speed, $W < 1$, maintaining rigid rotation will eventually lead to a
balance between the outward centrifugal force from rotation
and the inward force of gravity,
\beq
\frac{v_{\phi}^{2} (R_{\rm{K}})}{R_{\rm{K}}} = \frac{GM}{R_{\rm{K}}^{2}}
\, .
\eeq
Combining this with eqns. (\ref{wdef}) and (\ref{vprig}) gives a
simple expression for the associated
``Kepler radius'',
\beq
R_{\rm{K}} = W^{-2/3} \Rstar \, .
\label{rkdef}
\eeq
Unsupported material at radii $r < R_{\rm{K}}$ will tend to fall back toward the
star, but any material maintained in rigid-rotation to radii
$r > R_{\rm{K}}$ will have a centrifugal force that {\em exceeds} gravity, and
so will tend to be propelled further outward.
Indeed, any corotating material above an ``escape radius'', which is
only slightly beyond the Kepler radius,
\beq
R_{E} = 2^{1/3}\, R_{\rm{K}} \, ,
\eeq
will have sufficient rotational energy to escape altogether the local
gravitational potential, unless, of course, temporarily held down by the
magnetic field.

\subsection{2-D parameter grid of models}

\begin{figure}
\begin{center}
\vfill
\includegraphics[scale=0.5]{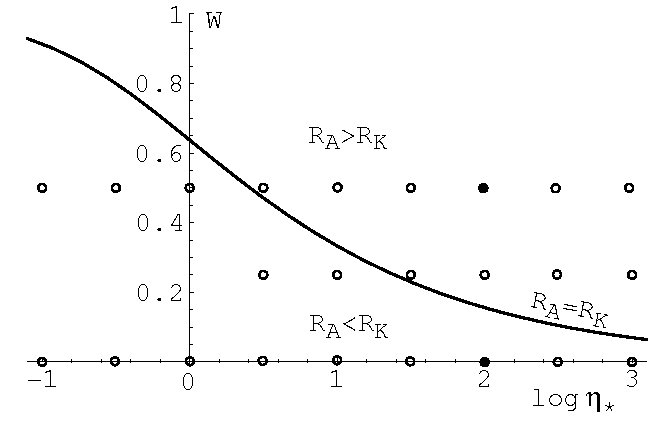}
\caption
{Plot of rotation vs. confinement parameter,
$W$ vs.\ $\log{\eta_\ast}$, to define the models in our
2D parameter study, represented here as filled and open circles,
with the filled circles representing models of principal focus for
more detailed analyses.
The solid curve respresents the contour for models with $R_{\rm{A}}=R_{\rm{K}}$
when the magnetic index is set to the dipole value $q=3$.
Models above this have $R_{\rm{A}} > R_{\rm{K}}$ and thus strong confinement
plus rapid rotation, whereas models below have $R_{\rm{A}} < R_{\rm{K}}$ and thus
either weak confinement or relatively slow rotation.
\label {param-space}}
\end{center}
\end{figure}

The circles in figure \ref{param-space} lay out the 2-D grid of models computed
for the present study, plotted in the plane of rotation parameter $W$
vs. log of the magnetic confinement parameter $\eta_{\ast}$.
The models along the x-axis include several specific cases already
examined in the non-rotating study in Paper I, with now however some
additional extensions toward the strong confinement limit,
viz. $\log \eta_{\ast}$~=~1.5, 2, 2.5 and 3.
In addition, there are now two new sets of corresponding models
with rotation parameters $W=$~0.25 and 0.5.
As noted, we do not consider faster rotation than $W=0.5$ because this
would introduce a significant stellar oblateness that would complicate
specification of the lower boundary condition for the spherical
coordinate system used in the ZEUS MHD code.

The solid curve in figure \ref{param-space} represents the parameter
combination for which  $R_{\rm{A}} = R_{\rm{K}}$ in the dipole case ($q=3$)
with velocity index $\beta=1$.
This  contour thus roughly divides the
parameter space diagonally:
models below and to the left have
only slow rotation and/or weak confinement, and so
$R_{\rm{A}} < R_{\rm{K}}$;
models above and to the right have fast rotation and/or strong confinement,
and so $R_{\rm{A}} > R_{\rm{K}}$.

Our analysis of the associated simulations show
that the lower-left models give generally quite similar
overall structure to what was found for the non-rotating models in
Paper I.
The more interesting cases are those
in the regions above and/or to the right, and
in the transition region with
$R_{\rm{A}} \approx R_{\rm{K}}$.

The transition region represents cases for which the magnetic spin-up is
just adequate to propel material into Keplerian orbit.
As such, it might seem to be appropriately fine-tuned to produce the
kind of magnetically torqued disk advocated by \citet{Cas2002}.
However, as discussed below and in previous papers
\citep{udD2006,Owo2006,Owo2006a,Owo2006b},
our simulations indicate that even for these optimal parameter cases,
the rotating magnetophere is characterized by a combination of infall
and outflow respectively below and above the Kepler radius, with no apparent
tendency to form an extended, stable, {\em Keplerian} disk.

On the other hand, in the limit of strong confinement with $R_{\rm{A}} \gg R_{\rm{K}}$,
the dominance of the field can confine the material in a
{\em rigid-body disk}, as postulated in the
``Rigidly Rotating Magnetosphere'' (RRM) formalism developed by
TO-05.
The full MHD simulations here allow us to directly test this RRM concept,
and define its limitations as wind material accumulates in the disk,
leading eventually to a centrifugally driven breakout overcoming the
confining magnetic tension
\citep[see][and Appendix of TO-05]{udD2006}.
Toward this goal, the magnetic confinement parameters considered here
extend to values ($\eta_{\ast} = 1000$) that significantly exceed the
maximum ($\eta_{\ast} = 10$) attempted in the non-rotating study of Paper I.

Such models with strong magnetic confinement are, in fact,
significantly more computationally challenging, since the
greater rigidity of the magnetic field implies a higher Alfv\'{e}n speed
(see Appendix),
and thus requires a smaller numerical time step to maintain stability
under the Courant criterion.
Indeed, this problem is often exacerbated by the tendency for the
strong, nearly horizontal field near the magnetic equator
to completely inhibit any wind base outflow there;
this leads then to short-lived nearly evacuated regions where the Alfv\'{e}n speed can become
exceedingly large, formally even approaching the speed of light!
To keep the time-step from becoming too small, we thus choose to
artificially add mass to these small evacuated regions at a level that is
sufficient to limit the local Alfv\'{e}n speed to
\beq
V_{A} \le \max(20,000~\rm{km/s}, \max(V_{Ap}))
\, ,
\label{vamax}
\eeq
where
\beq
\max(V_{Ap}) = 0.65 \sqrt{\eta_{\ast}}\,  \vinf
\, ,
\label{vapmax}
\eeq
is the expected maximum polar Alfv\'{e}n speed,
as given by the analysis in the Appendix.
We check that the amount of artificially added mass
is still quite insignificant compared
to the global mass loss in the wind,
i.e.\ less than a percent in even the most extreme ($\eta_{\ast}=10^{3}$)
cases.

Moreover, the larger Alfv\'{e}n radius means such models need generally
a larger outer boundary radius, and the larger breakout timescale
(as predicted in the Appendix of TO-05) means that
models have to be run longer to cover the breakout cycles and
associated accumulation of mass in any RRM disk.
Finally, as discussed further below,
the closed magnetic topology of episodic outbursts
can complicate the proper specification of the outer boundary
condition, and in practice reflection effects as these outbursts
are advected through the boundary can occasionally even
halt the computation altogether.
In summary, the extension of MHD simulations into the very strong
confinement domain remains an ongoing challenge.

\subsection{Stellar and wind parameters}

Much of the procedures in the current study follows Paper I.
Specifically, we use the same standard non-magnetic and {\em non-rotating} wind
model used in Paper 1, but now at the initial time we suddenly
introduce both a dipole magnetic field, {\em and a surface rotation} at the
lower boundary, with both defined relative to a common polar axis.
This standard model has
stellar parameters representative of a typical OB supergiant,
with  a radius $R_{\ast} = 19 \Rsun$, a luminosity $L = 10^6 \Lsun$,
and an effective mass of $M= 25\Msun$.
(This reflects a factor two reduction below the Newtonian mass to  account
for the outward force from the electron scattering continuum.)
This radius and mass imply an effective equatorial surface orbital speed of
$V_{\rm{orb}} \approx $~500~km/s.

The line-driving assumes a CAK power index $\alpha=0.6$ and a line normalization such
that the non-magnetic, non-rotating wind has a mass loss rate of about
${\dot M} \approx 3 \times 10^{-6} M_{\odot}$/yr and a terminal
speed of about $\vinf \approx 2400$~km/s.
Since both the stellar and wind parameters are fixed,
we vary the magnetic confinement parameter $\eta_\ast$ solely through
the variations in the assumed equatorial surface field strength, $B_{eq}$.
As noted above, we do not consider rotation parameters $W > 1/2$
since this would deform the stellar surface and  require
consideration of gravity darkening,
neither of which are taken into account in our models.

One further difference compared to the simulations in Paper I is that we
find it necessary to run the rotating models here for a longer time in
order to identify properties of a relaxed, quasi-stationary asymptotic
state (especially for the strong confinement models).
To facilitate comparison among different cases, we standardize a run to duration
of $t=0$ to 3~Msec, which is already 6 times longer than the
0.5~Msec used in Paper I.
But we have also run selected models for a longer time,
e.g. 6~Msec for our standard case with $W=1/2$ and $\eta_{\ast}=100$.
Required run times per model are typically about one to two weeks on a standard
workstation.

\begin{figure*}
\begin{center}
\includegraphics[scale=.65]{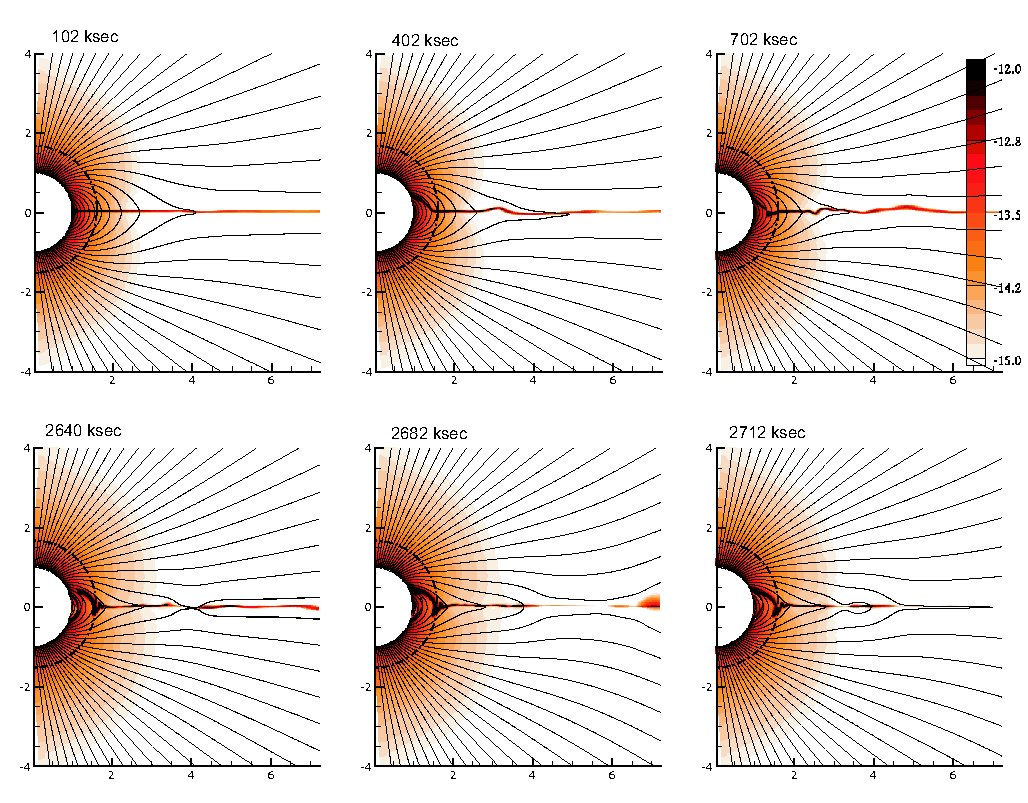}
\caption{
Snapshots of density (in cgs units on a logarithmic color-scale) and
field lines (solid lines) at the labeled time intervals for the model
with $\eta_\ast=100$ and $W=1/2$.
The top panels show the model during a relatively quiescent period,
when a dense rigidly rotating disk is being gradually built up.
Note, however that material below the Kepler co-rotation radius
($R_K$, shown as a dashed circle)
falls back onto the stellar surface, due to the lack of
sufficient centrifugal support.
The extent of the disk in this phase
is determined by the magnetic field strength and extends
up to the Alfv\'{e}n radius $R_{\rm{A}} \approx 3.4 R_{\ast}$,
which is somewhat above the maximum outer radius of closed magnetic
loops, $R_{\rm{c}}$.
The bottom panels show the model later in the evolution, during
one of the episodic centrifugal breakout events.
}
\label {logd-eta100-fig}
\end{center}
\end{figure*}

\begin{figure*}
\begin{center}
\includegraphics[scale=.65]{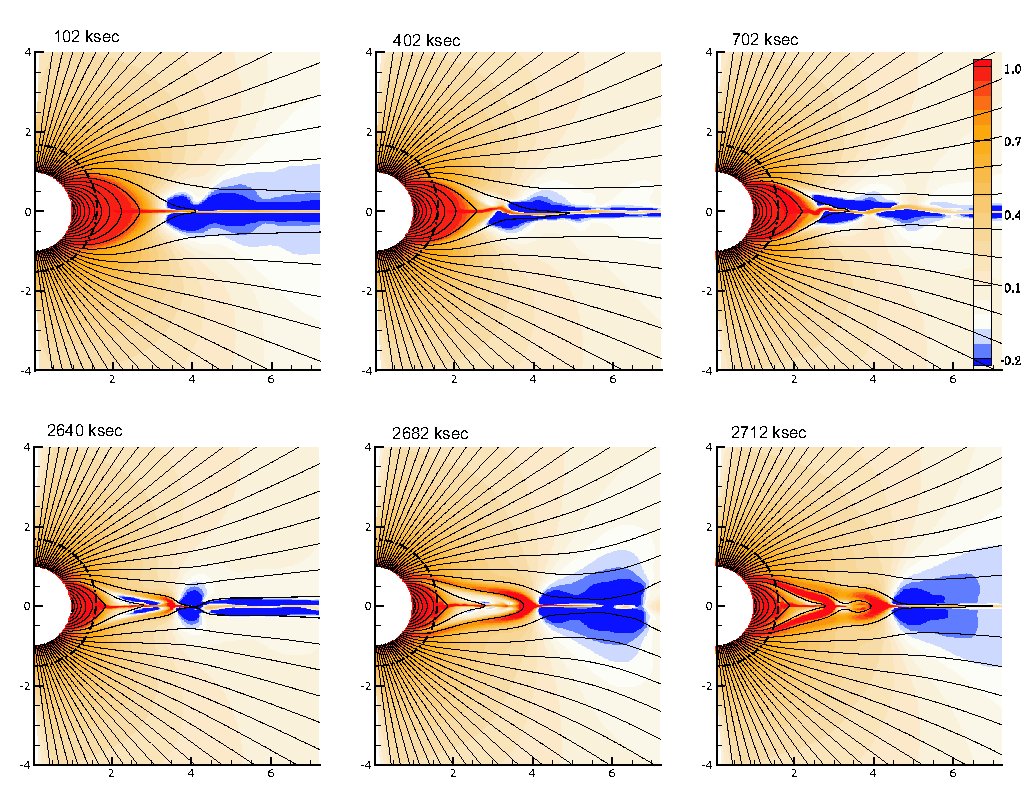}
\caption {
For the same model and same time snapshots as in figure \ref{logd-eta100-fig},
the azimuthal speed scaled by the local co-rotation speed,
$\chi \equiv v_\phi/\Omega r \sin {\theta}$) .
During the quiescent period the magnetosphere extends nearly to the
closure radius $R_{\rm{c}} \approx 2.7 \Rstar$,
and co-rotates almost rigidly with the star (top panels).
However, during episodic centrifugal breakouts (bottom panels)
this rigidly rotating magnetosphere shrinks nearly to $R_K \approx 1.6
\Rstar$, represented by the dashed circle.
In addition to the mass lost to the breakout, this also helps some of
the disk mass to leak inward as infall back onto the star.
}
\label{chi-eta100-fig}
\end{center}
\end{figure*}

\section{RESULTS}

\subsection{Co-rotation and rigid disk}

As a standard example to frame the overall study here,
let us first focus on this case with confinement $\eta_\ast= 100$
and rotation  $W=1/2$ ($V_{\rm{rot}} = 250$~km/s).
Figures \ref{logd-eta100-fig}  show a series
of time snapshots of the 2D spatial configuration of the magnetic field
(solid lines), with the color scale representing logarithm of density;
figure \ref{chi-eta100-fig} give a similar time sequence for the
azimuthal flow speed, scaled relative to the value that would
occur in rigid-rotation, i.e.
$\chi \equiv v_\phi(r)/\Omega r \sin \theta$,
where $\Omega \equiv V_{\rm{rot}}/R_{\ast}$
is the star's angular rotation  frequency.
The time snapshots were chosen to illustrate both relatively quiescent
intervals (top panels),
and phases with dynamic centrifugal breakout (bottom panels).
The dashed circle represents the Kepler co-rotation radius
at the equator ($R_K \approx 1.6 \Rstar$).

In the evolution immediately following the initial condition,
the magnetic field channels wind material toward the tops
of closed loops near the equator, where the collision with the opposite
stream leads to a dense disk-like structure (see top panels).
But the gas is also generally torqued by the field,
with, as can be seen in the upper panels of figure \ref{chi-eta100-fig},
material in the closed magnetosphere up to
$R_{\rm{c}} \approx 2.7 R_{\ast}$
kept nearly in rigid-body co-rotation with the star.
Note that these closed, rigidly rotating loops thus extend through
and beyond the Kepler radius.
For any material trapped on loops below $R_{\rm{K}}$, the outward centrifugal support
is less than the inward pull of gravity;
since much of this material is compressed into clumps that are too dense to be
significantly line-driven, it thus eventually falls back to the star
following
complex patterns along the closed-field loops.

By contrast, the dense material {\it above} the dashed line at $R_{\rm{K}}$ has a
net radially {\em outward} force from the centrifugal acceleration vs.\ gravity.
Still, during the initial build-up of this material at the tops of
loops above $R_{\rm{K}}$, the magnetic field provides tension
that is strong enough to hold it down, forming then a segment of the
{\em rigidly rotating disk} predicted in the analytic RRM analysis by
TO-05.
However, much as anticipated in the Appendix of their paper, eventually
material in the outer region of this RRM accumulates to sufficient
density to force open the magnetic field, leading to the kind of
centrifugally driven breakout events simulated in \citep{udD2006}.
This is illustrated here in the bottom panels of figure \ref{logd-eta100-fig}.

Note however from figure \ref{chi-eta100-fig} that certain regions, marked in
blue, actually have a net azimuthal motion that is {\it against} the sense of
the stellar rotation.
This surprising and counterintuitive result is not a numerical artifact, but
rather is related to a reverse torque effect that occurs in regions of
rapid wind acceleration.
This was first discussed by \citet*{MacFri1987}, who extended
the classic \citet{WebDav1967} 1-D magnetic monopole
rotation model for the solar wind to the case of the more rapidly
accelerating line-driven winds.
The 2-D analog for the dipole field here has little impact on our study of
rotation and magnetic confinement, and so we defer further discussion to
an upcoming paper that focuses on the role of the magnetic field in
outward angular momentum transport and stellar spindown.

{
The centrifugal breakouts occuring in these simulations necessarily imply a
breakdown in the basic formulation for ideal MHD within the Zeus code.
In equatorial regions where the wind or centrifugal terms
stretches out field lines of opposite polarity,
the finite grid resolution allows effective reconnection, with the
associated release of magnetic energy effectively lost instantaneously
(e.g. due to radiative cooling) in these isothermal simulations.
This is admittedly a simplified representation of the
very complex physics thought to occur in actual reconnection, which
indeed is an intense area of modern plasma physics research \citep{Shay1999}.
But in the present context of {\em driven} reconnection,
the overall global evolution seems likely to be set by the central
competition between magnetic confinement and centrifugal breakout,
with relatively little sensitivity to the details of local
reconnection sites.
}

\subsection{Global evolution of equatorial disk in radius and time}

\begin{figure}
\begin{center}
\vfill
\includegraphics[scale=.5]{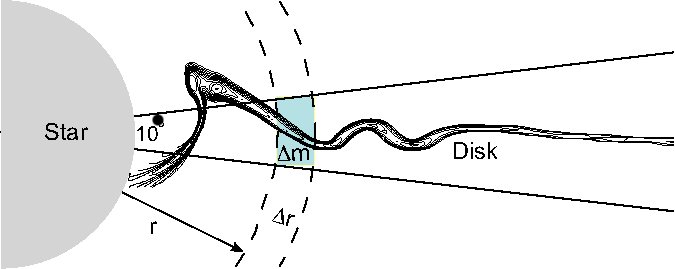}
\caption
{
Schematic diagram illustrating the computation of
the radial mass distribution of the equatorial disk,
$dm_{e}/dr$ (see eqn. \ref{dmedrDef}).
We choose a cone angle of $\Delta \theta = 10^{o}$ centered
on the magnetic equator to
encompass most of the material in the disk.
We then compute the total mass $\Delta m$ contained in a narrow strip
$\Delta r$ within the cone.
Note, however, that some of the material may not lie within the
cone during infall episodes.
\label{dmdr-sketch}
}
\end{center}
\end{figure}

{
A key result of the simulations here is that there is really
{\em no true steady state} possible, since the secular buildup of
material in the disk must eventually lead to an episodic material breakout
once the centrifugal forces overwhelm the finite magnetic tension.
}
One primary goal of the more extensive parameter study here is to examine in
detail the nature of this build-up and dissipation of mass in an
RRM disk, and how this varies with the changes in the rotation rate
and magnetic confinement.
To facilitate illustration of these competing processes, let us define
a radial mass distribution of the disk, computed at each radius $r$
in terms of the mass within some specified
co-latitude range about the equator,
\beq
\frac{dm_{\rm{e}} (r,t)}{dr} \equiv 2 \pi r^{2}
\int_{\pi/2-\Delta \theta/2}^{\pi/2+\Delta \theta/2} \, \rho(r,\theta,t) \,
\sin \theta \, d\theta
\, .
\label{dmedrDef}
\eeq
To isolate the disk but not miss too much disk material during various
oscillations about the equator, we choose a narrow, but not-too-limited
range $\Delta \theta = 10^{o}$.
Figure \ref{dmdr-sketch} shows schematically how this is computed.

\begin{figure}
\begin{center}
\vfill
\includegraphics[scale=0.23]{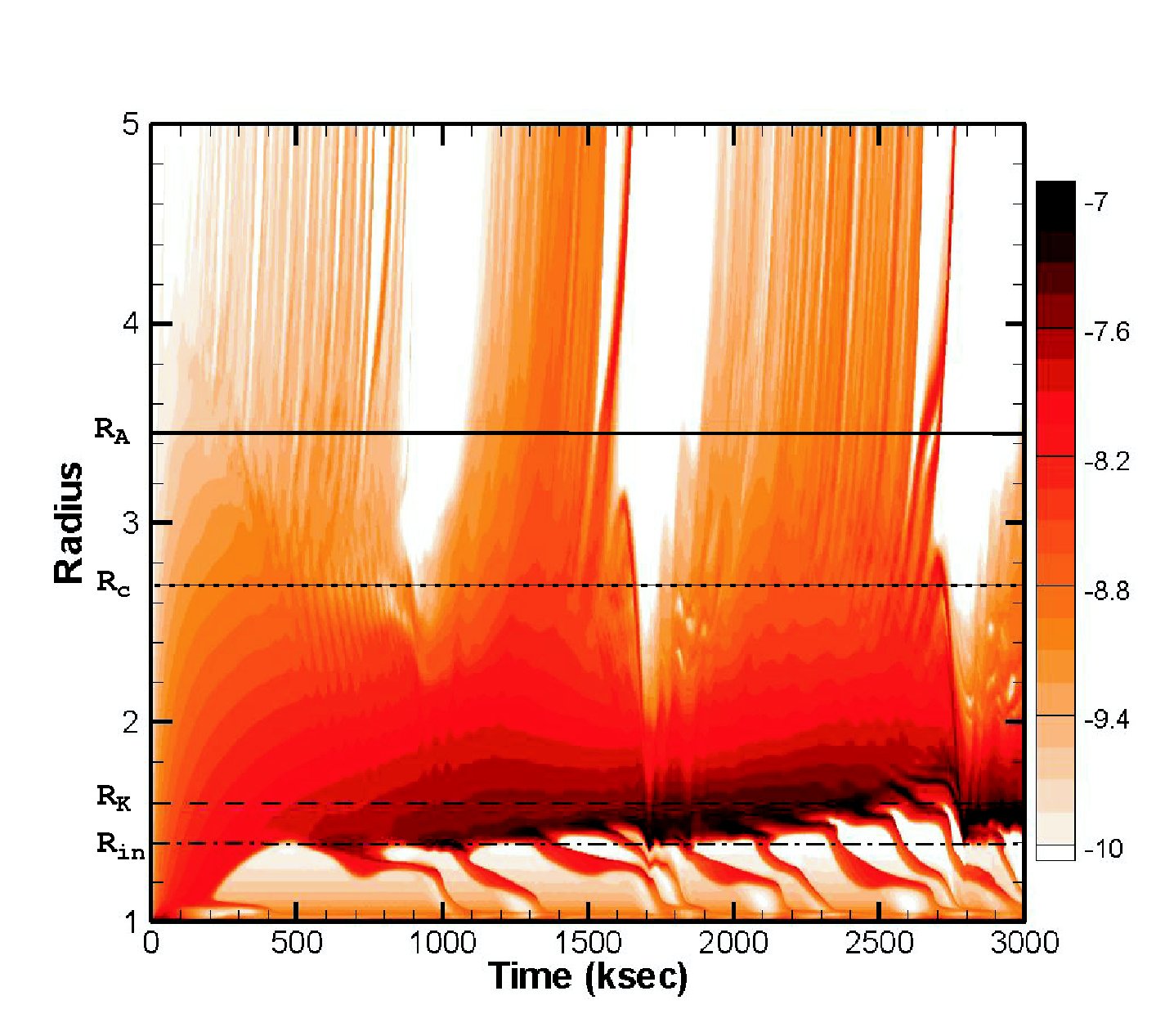}
\caption
{
For MHD simulations of the standard model case with $\eta_{\ast}=100$ and
$W=1/2$, the logarithm of the radial distribution of mass, $dm_{e}/dr$, within
a cone of width $\Delta \theta = 10^{o}$ centered on the equator,
plotted vs. time and radius (in units of stellar radius).
The color bar shows $\log(dm_{e}/dr)$ in units of  $M_{\odot}/\Rstar$.
The horizontal lines indicate the Alfv\'{e}n radius $R_{\rm{A}}$ (solid),
the maximum loop closure radius $R_{\rm{c}}$ (dotted),
the Kepler radius $R_{\rm{K}}$ (dashed),
and the inner RRM disk radius $R_{in} = (2/3)^{1/3} R_{\rm{K}}$
(dot-dashed).
Note the accumulation of mass in a disk near the Kepler radius, and
how this is limited  both
by relatively frequent (ca. every 200~ksec)
intervals of downward infall,
and by less-frequent (ca. every 1~Msec) major disruptions with
substantial upward breakout.
\label{dmdr-vs-rt}
}
\end{center}
\end{figure}

For the same standard case ($W=0.5$, $\eta_{\ast}=100$)
shown in figures \ref{logd-eta100-fig} and
\ref{chi-eta100-fig}, figure \ref{dmdr-vs-rt}
shows a colorscale plot of this disk mass
vs. radius (on the ordinate) and time (on the abscissa).
The horizontal  lines mark, from top to bottom,
the estimated Alfv\'{e}n radius $R_{\rm{A}} \approx 3.4 \Rstar$,
the loop closure radius $R_{\rm{c}}$,
Kepler radius $R_{\rm{K}} \approx 1.6 \Rstar$,
and inner disk radius,
$R_{in} \equiv (2/3)^{1/3} R_{\rm{K}} \approx 1.4 \Rstar$
[see eqn. (19) of TO-05].
Within the RRM model,
the last represents the location where the effective potential along
a rigid-field loop first develops a local minimum, which can then trap material
fed from the wind.

The plot shows quite succinctly, and vividly, the global time
evolution of the equatorial disk material.
Initially mass builds up in the region around
$R_{\rm{K}}$, but then there appear repeated episodes of infall of inner
disk material back onto the star, about every 200~ksec or so.
This leads to a gradual outward progression to the lower edge of the disk
material.

But over a somewhat longer timescale, about every 1~Msec or so,
there appears another,  somewhat different kind of disruption,
one that starts higher up, closer to the closure and Alfv\'{e}n radii.
This is characterized by outward ejection of the upper disk mass, but
then also a ``rebound'' that propagates back down toward the Kepler
radius, pushing the trapped disk material inward, and inducing a
further leakage of disk mass through downward infall.
The overall effect is to regulate the disk mass so that, by the end of
the simulation at $t=3$~Msec, the addition of new
material from the wind becomes roughly balanced by the losses to both
infall and ejection.

The overall structure is certainly quite dynamic, but near the Kepler
radius there nonetheless appears to be a quasi-permanent disk segment that
corresponds roughly to what is predicted by the RRM analytic
analysis of TO-05, as well as by the recent time-dependent
Rigid-Field Hydrodynamics (RFHD)
simulations of Townsend, Owocki, and ud-Doula (2007).

\subsection{Comparison with RRM breakout analysis}

Both the RRM and RFHD approaches are based on the idealization that
the field is arbitrarily strong, and so remains perfectly rigid
regardless of the amount of material in either the wind outflow or
the disk buildup.
But the associated discussion for both approaches recognized that the
secular accumulation of material in the rigid-body disk would
eventually cause the outward centrifugal forces to overwhelm the
available inward tension associated with any large but finite magnetic field.
In fact, the Appendix of TO-05 presents a simplified, but quantitative
analysis of the resulting expected ``breakout'' of accumulated disk
material.
This anticipates, at least in general terms, several aspects of the
processes seen in the present MHD simulations.
In particular, it makes quite specific predictions for both the
breakout timescales as a function of radius, and for the asymptotic
mass accumulation near the Kepler radius.
In this section, let us thus attempt a specific, semi-quantitative
comparison between those predictions and the results of the
MHD simulations.

\subsubsection{Breakout timescale}

Without benefit of the {\em global} dynamical picture available from the
MHD simulations here, the TO-05 breakout analysis focused instead on the
conditions for breakout at each {\em local} radius, conveniently scaled
in terms of the Kepler radius as $\xi \equiv r/R_{\rm{K}}$.
From TO-05 eqn.\ (A6) we find that,
in terms of the free-fall time $t_{\rm {ff}} \equiv \Rstar/V_{\rm{esc}}$,
the breakout time for some scaled outer disk radius
$\xi_{\rm {o}}$ is given by
\begin{equation}
t_{\rm{b}} \approx
\eta_{\ast} t_{\rm{ff}}  \, \frac{6 \xi_{\ast}/\xi_{\rm {o}} }{{\xi_{\rm {o}}}^{3}-1} \, .
\label{eqn:time-b}
\end{equation}
Here we have approximated $\sqrt{\pi}/\mu_{\ast} \approx 2$
(cf. eqn.\ A8 of TO-05),
and used the ratio of wind terminal speed to escape speed,
$\vinf/V_{\rm{esc}} \approx 3$, to convert the {\em disk} confinement
parameter (also denoted $\eta_{\ast}$) in TO-05 into the {\em wind}
confinement parameter defined here\footnote{
Although a footnote in the Appendix of TO-05
seems to imply that $B_{\ast}$ and
$B_{eq}$ are distinct, they are in fact both equal to the field
strength at the equatorial surface.}.
Note also that the
Kepler-scaled stellar radius can be written as
$\xi_{\ast} =\Rstar/R_{\rm{K}} = W^{2/3}$.

Applying our stellar free-fall time $t_{\rm{ff}} \approx 19$~ksec,
then for our standard ($\eta_{\ast}=100$, $W=1/2$) model,
the predicted breakout time is
\beq
t_{\rm {b}} \approx \frac{7 {\rm Msec}}{\xi_{\rm {o}}({\xi_{\rm {o}}}^{3}-1)}
\, .
\eeq
In the MHD simulations for this standard case, the breakouts
seem to originate around $R_{\rm{c}} \approx 2.7 \Rstar$,
for example as indicated in figure \ref{dmdr-vs-rt} by the ``bi-furcations''
between  upward and downward  tracks that start at $r \ltwig 3 \Rstar$
for times around 800, 1600, and 2700~ksec.
If we thus approximate the outer disk radius by this maximum loop closure
radius, we find $\xi_{\rm {o}} = R_{\rm{c}}/R_{\rm{K}}  \approx 1.7$ and
so $t_{\rm{b}} \approx 1$~Msec, about the timescale between major breakout
eruptions seen in these same MHD simulations.

The TO-05 breakout analysis envisioned a hierarchy of breakout
timescales, with more frequent eruptions occurring at larger radii;
but its concluding paragraph also anticipated (partly based on early
versions of the MHD simulations described here) that breakouts
originating within $r \ltwig 2 R_{\rm{K}}$ could also lead to substantial
disruption of the entire magnetosphere.
The simulations here do indeed show such major disruptions, but even
after these there remains substantial mass near the Kepler radius.

\begin{figure}
\begin{center}
\vfill
\includegraphics[scale=0.85]{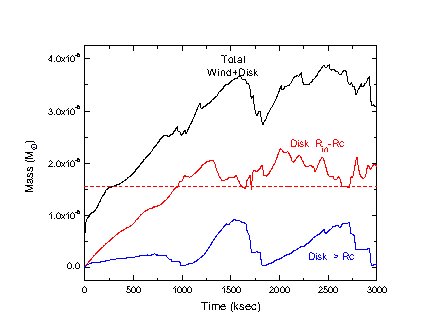}
\caption
{Cumulative mass vs. time in MHD simulations for the standard model
case.
The top curve shows the total mass in the entire grid, i.e.  including
both disk and wind material.
The middle curve shows the disk mass within $5^{o}$ of the equator,
and within radii bounded by the inner RRM disk radius $R_{in}$ and
an outer radius $R_{\rm {o}} = R_{\rm{c}}$.
The bottom curve shows equatorial mass above this outer radius.
The horizontal dashed line indicates the cumulative disk mass
predicted by the RRM breakout analysis, using this same outer radius.
Note that it matches quite well the varying
asymptotic value of the middle curve,
showing the disk mass in the MHD simulation.
\label{m-vs-t}
}
\end{center}
\end{figure}

\subsubsection{Accumulated disk mass}

For a disk with a scaled outer breakout radius $\xi_{\rm {o}}$, eqn. (A10)
of TO-05 predicts a specific scaling for the total asymptotic disk
mass, which in terms of the parameters here can be written,
\begin{eqnarray}
m_{\rm{d}} (\xi_{\rm {o}} ) &=&
\frac{ 3 \sqrt{\pi} \,{\dot M} \, t_{\rm{ff}} \, \eta_{\ast} \, W^{4/3}}
{{\xi_{\rm {o}}}^{2} ({\xi_{\rm {o}}}^{2} + \xi_{\rm {o}} + 1)}
\nonumber
\\
&=& 2.1 \times 10^{-9} M_{\odot} \,
\frac{\eta_{\ast} W^{4/3}}
{{\xi_{\rm {o}}}^{2} ({\xi_{\rm {o}}}^{2} + \xi_{\rm {o}} + 1)}
\, ,
\end{eqnarray}
where the latter gives the numerical scaling for the stellar and wind
parameters used here.
If we then apply the confinement and parameters of our standard model,
and use the characteristic breakout radius $\xi_{\rm {o}} \approx 1.7$
adopted above, we obtain a predicted total disk mass of
$m_{\rm{d}} \approx 1.6 \times 10^{-8} M_{\odot}$.

Figure \ref{m-vs-t} compares
this predicted mass (red horizontal dashed curve) with
the time variation of three types of cumulative
mass in the standard model MHD results.
Specifically, the top curve (black) shows the total integrated
mass in the entire grid, including regions of both wind and disk;
the middle curve (red) gives the cumulative mass in the equatorial disk
region between the inner radius
$R_{in} \approx 0.87 R_{\rm{K}} \approx 1.4 \Rstar$ and outer radius
$R_{\rm {o}} \approx R_{\rm{c}} \approx 2.7 \Rstar$;
and finally, the bottom curve (blue) shows the equatorial mass above
this outer disk radius, $r > R_{\rm {o}} \approx R_{\rm{c}}$.

Note in particular that the predicted asymptotic disk mass agrees
remarkably well with this MHD simulation disk mass as shown by the
middle, red curve.
Of course, this is partly fortuitous, since just a 10\% change
in the choice of outer disk radius $\xi_{\rm {o}}$ would imply a ca. 40\% change
in disk mass.
But the overall, order-magnitude agreement seems likely to be quite
robust, and so provides a nice consistency check for both the RRM analysis
and numerical simulations.

Comparison of the middle and upper curves in figure \ref{m-vs-t}
further shows that the total mass in the thin, radially limited disk
represents about half of the total mass in the entire model.
In part, this reflects the fact that, for such a strong confinement model,
a large fraction of the outgoing wind gets channeled into the disk.
But another factor is that the wind material is flowing outward at a
very high speed, and so has a much shorter ``residency time'' than the
trapped, relatively static material in the disk.

\subsubsection{Limitations  of a localized breakout description}

Despite this general success of the local breakout analysis in matching
both the overall breakout timescale and the accumulated disk mass of
this MHD simulation, the specifics of the dynamical evolution seen in the
simulation make clear that the breakout process is really a {\em
global} phenomenon.
As the accumulation of material in the outer disk regions stresses and
eventually overcomes the inward restraint of the magnetic field, the
associated outward stretching alters the global field, including in
the inner regions near the Kepler radius.
Moreover, once a breakout occurs, the release of this stretching
causes the inner, closed field lines to snap back inward, much like a
stretched rubber band after release.
The overshoot can push disk material {\em below} the Kepler radius
and trigger infall back onto the star.

Overall, the wind-fed accumulation of disk mass is thus balanced not
just by ejection outward, but also by infall inward.
In contrast to the idealized picture of the breakout analysis, which
formally predicts the timescale for breakout (and thus emptying) of
material right at the Kepler radius to become arbitrarily long, the
dynamic oscillation and associated inward spillage of material limits
the asymptotic mass accumulated in this region.
This new perspective on the dynamical nature of the disk mass budget
has potentially important implications for modelling and interpreting
observational diagnostics ( see \citealt{Tow2005}; \citealt*{Tow2007}).


\subsection{Comparison with non-rotating model}

\begin{figure}
\begin{center}
\vfill
\includegraphics[scale=0.25]{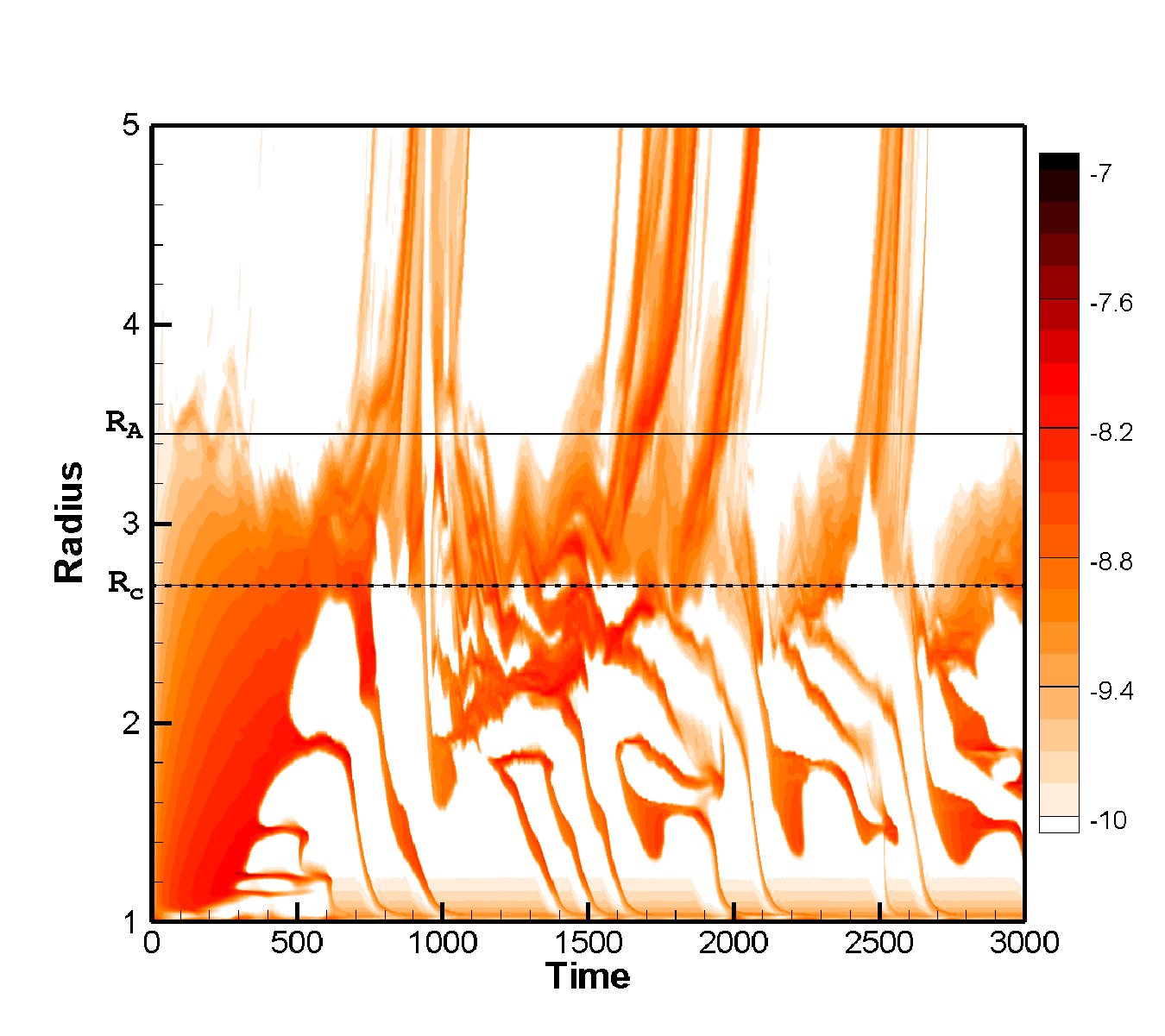}
\caption{
As in figure \ref{dmdr-vs-rt}, the logarithm of the radial distribution of equatorial mass,
$dm_{e}/dr$, again plotted vs. time and radius for a strong confinement
($\eta_{\ast}=100$) model, but now with {\em no rotation} ($W=0$).
The horizontal lines indicate the Alfv\'{e}n radius $R_{\rm{A}}$ (solid)
and loop closure radius $R_{\rm{c}}$ (dotted).
Note there are again complex patterns of breakout and infall, but now
with more irregular timescales, and with the infall extending up
to regions of breakout, near the loop closure radius.
As such, there is no longer an extended region of mass accumulation
into a circumstellar disk.
\label{dmdr-vs-rt-w0}
}
\end{center}
\end{figure}

To demonstrate further the role of rotation in how magnetic
fields influence a wind outflow, let us now compare the results of
this $\eta_{\ast}=100$, $W=1/2$ case with the corresponding
non-rotating model.
Figure \ref{dmdr-vs-rt-w0} illustrates the dynamic evolution of
equatorial mass for this strong-confinement case without rotation.
Comparison with figure \ref{dmdr-vs-rt} shows that there are still both
breakout and infall episodes, but now with both originating from nearly
the same location, at roughly the loop closure radius
$R_{\rm{c}} $.
This infall from throughout most of the closed field region reflects
the lack of any centrifugal support against gravity, and as such,
there is no longer any accumulation of material into a circumstellar
disk.
The breakouts remain, but instead of being driven by centrifugal forces,
these are now the result of entrainment of the field with the
outflowing wind.
The timescales for both breakout and infall are comparable to the
rotating case, but seem somewhat more irregular.
Overall, without the buildup in the disk, there is significantly less
mass in the magnetosphere than in the rotating case.

\begin{figure}
\begin{center}
\vfill \includegraphics [scale=1.00]{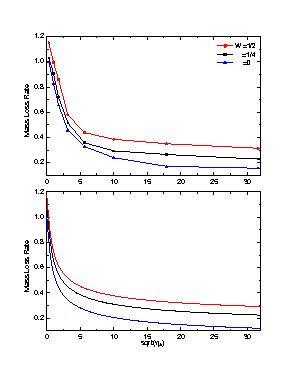}
\caption{
Top panel:
Average mass loss rate (in units of the non-rotating, non-magnetic case) for MHD models
with various rotational rates
$W=0, 1/4, 1/2 $ plotted against $\sqrt {\eta_\ast} \propto B_\ast$.
Note that for $W=1/2$ the mass loss rate is nearly constant for models
with high confinement $\eta_\ast >25$.
Bottom panel: Corresponding mass loss rates from the analytic
scaling formula in eqn. (\ref{mdred}).
\label {mdot-fig} }
\end{center}
\end{figure}

\subsection{Results for variation in parameters}

To provide a wider context to these detailed results for
specific cases with strong confinement, let us now
analyze results spanning a broader range of the 2-D parameter
space for rotation and confinement strength.
Before focusing further on equatorial disk material,
let us first briefly consider the effects of various magnetic field
strengths  and rotation rates on the {\em global}
mass loss from the stellar wind.

\subsubsection{Global mass loss}

As discussed in Paper I, one general effect of a strong magnetic field
on a wind is the confinement and inhibition of the outflow within a belt
around the magnetic equator.
For a dipole field line that reaches up to the maximum closure radius
$R_{\rm{c}}$, the  co-latitude $\theta_{c}$ at the surface footpoint satisfies
\beq
\sin \theta_{\rm {c}} = \sqrt{\Rstar/R_{\rm{c}}}
\, ,
\eeq
The fraction of the stellar surface area that is covered
by closed field lines is given by $\cos \theta_{\rm{c}}$,
leaving thus only the remaining fraction $1- \cos \theta_{\rm{c}}$
as the source of wind mass loss.
For the non-rotating case,
we can thus use this open-field fraction to estimate the overall magnetic
reduction in the mass loss rate,
\beq
\frac{{\dot M_{\rm {B}}}}{{\dot M_{\rm {B}=0}}} \approx 1 - \sqrt{1-\Rstar/R_{\rm{c}}}
\label{mdredw0}
\, ,
\eeq
where $R_{\rm{c}}$ is evaluated from eqn.\ (\ref{rcdef}), with
the Alfv\'{e}n radius $R_{\rm{A}}$ from eqn.\ (\ref{raapp}) using
$q=3$.
Note that this ignores higher order effects, such as the reduction of
the mass flux in open field regions due to the tilt of surface field
relative to the radial direction for wind driving \citep[see][]{OwoudD2004}.

The two panels in figure \ref{mdot-fig} compare the mass loss rate vs.
$\sqrt{\eta_{\ast}}$ ($\propto B$) for both simple analytic scalings
(bottom) and results of numerical MHD simulations (top),
with the lower, middle, and upper curves in each panel
corresponding to the $W=0$, 1/4, and 1/2 rotation models.
For the non-rotating case, the numerical and analytic results
shown in the lower curves are in good overall agreement.
But for the rotating case, the upper curves in the top panel show
that the tendency of the strong field to reduce the overall
mass loss rate is somewhat compensated by faster rotation,
and in the $W=1/2$ case, even flattens to nearly constant toward
the limit of strong confinement.
This reflects the additional effect of centrifugal forces in driving
the breakout of material initially trapped in closed loops near and
below the confinement radius $R_{\rm{c}}$.
In effect, the rotation allows eventual breakout from loops that are
some factor times the Kepler radius, say $2 R_{\rm{K}}$.

To take this into account in an analytic scaling formula, the upper
two curves in the lower panel use a modified form of
eqn.\ (\ref{mdredw0}),
\beq
\frac{{\dot M_{\rm {B}}}}{{\dot M_{\rm{B}=0}}} \approx 1 - \sqrt{1-\Rstar/R_{\rm{c}}}
+ 1 - \sqrt{1-0.5*\Rstar/ R_{\rm{K}}}
\label{mdred}
\, ,
\eeq
which effectively sums separate contributions from polar opening and
rotational breakout, with the closure and Kepler radii computed from
eqns.\ (\ref{raapp}), (\ref{rcdef}), and (\ref{rkdef}).
With this generalized scaling, the overall variations of the
analytic curves in the lower panel roughly match the
corresponding MHD results in the upper panel.

\begin{figure*}
\begin{center}
\includegraphics [scale=0.64]{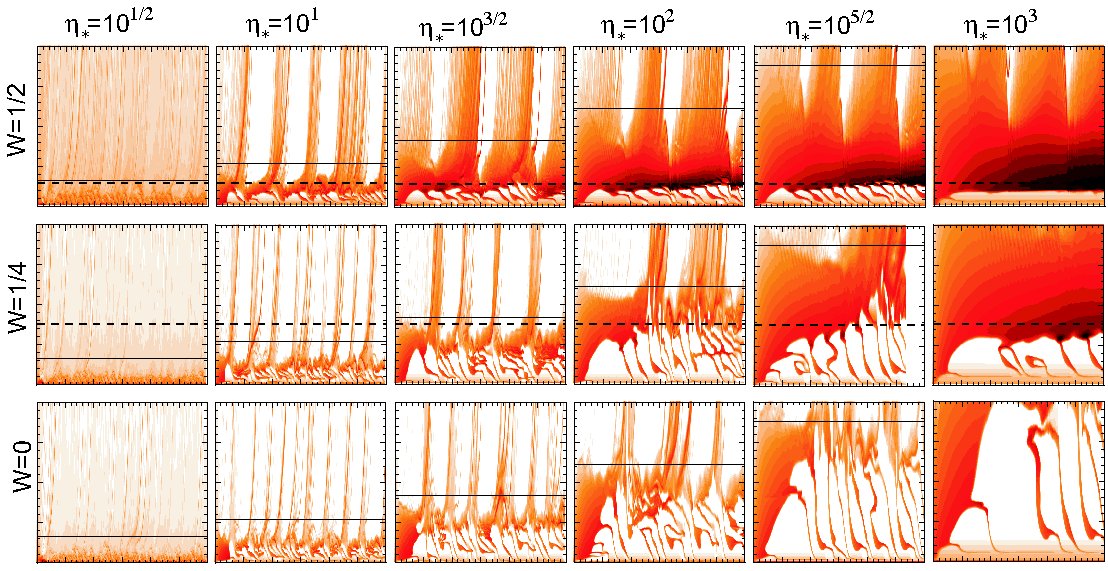}
\caption[] {
Logarithm of radial distribution of equatorial disk mass, $dm_{e}/dr$ vs. radius
and time, for a mosaic of models with magnetic confinements
$\log(\eta_{\ast})=$~1/2, 1, 3/2, 2, 5/2 and 3 (columns from left)
and rotations $W=$~0, 1/4, and 1/2 (rows from bottom) with
the same contour levels and the ranges for time and radius as in figure \ref{dmdr-vs-rt}.
Again, the horizontal lines indicate the Alfv\'{e}n radius $R_{\rm{A}}$ (solid)
and the Kepler radius $R_{\rm{K}}$ (dashed).
\label {dmdr-vs-params} }
\end{center}
\end{figure*}

\subsubsection{Equatorial mass and disk}

Let us next examine how the equatorial disk region is affected by
variations in the magnetic confinement and rotation parameters.
Figure \ref{dmdr-vs-params} compares the radius and time evolution of
the equatorial mass, $dm_{e}/dr$, for an mosaic of models with
various $\eta_{\ast}$ and $W$.
The comparison provides a global overview of how the equatorial mass
evolution is affected by changes in confinement and rotation.

For weak rotation and confinement cases in the lower left panels,
material generally escapes outward without much
infall, with only a modest rotational enhancement in mass loss.
But most other models again show a complex competition between infall and
breakout, with the latter always being less frequent and stronger.

In particular, this complex combination of infall and breakout
also dominates the $R_{\rm{A}} \approx R_{\rm{K}}$ transition models
identified in figure \ref{param-space}, i.e. the ones here with
$\log \eta_{\ast} = 1/2$ and $W=1/2$ or
$\log \eta_{\ast} = 3/2$ and $W=1/4$.
Such models might seem optimally fine-tuned to propel material into Keplerian
orbit, and yet they show no apparent tendency for material to accumulate into
the extended, Keplerian disk envisioned in the MTD scenario
suggested by \citet{Cas2002}.
The lack of a sharp outer cutoff in the large-scale dipole field makes it
incompatible with the shear of a Keplerian disk, and
without the closed loops that hold down a rigid disk in the
strong-confinement limit, material is propelled
outward to escape, rather than into a stable Keplerian orbit.

As expected, accumulation into such a rigid-body disk is strongest for the
fastest rotation, and strongest confinement, as shown by models at the
upper right.
For strong confinement but slow or no rotation, the material
infall comes from
a greater height, set by the closure radius, which increases
roughly as $R_{\rm{c}} \sim \eta_{\ast}^{1/4}$.

This larger infall height seems also to lead to a somewhat longer
infall timescale.
Likewise, the breakout timescale also seems to increase for models
with stronger confinement paramater $\eta_{\ast}$,
but not quite in the linear proportion that
might be suggested by eqn.\ (\ref{eqn:time-b}).
The reason is that the scaled outer confinement radius $\xi_{\rm {o}}$
appears in the denominator, and since this also increases with
confinement through it dependence on the closure the radius $R_{\rm{c}}$,
the net effect is to weaken the $\eta_{\ast}$ sensitivity of $t_{\rm{b}}$,
especially in the strong confinement limit $\eta_{\ast} \gg 1$.

Note also that the $W=1/2$ rotation model with the strongest confinement,
$\eta_{\ast} = 1000$, is relatively stable, without the repeated
equatorial infall events seen in other models.
Instead of the extensive north-south disk oscillations seen in other models,
in this case the variations of the equatorial disk are mostly symmetric about
the equator, and thus do not induce as much ``spillage'' back onto the star.
The recent analysis of  ``Rigid-Field Hydrodynamics'' (RFHD) models by
\citet{Tow2007} show that both types of oscillation modes are allowed,
with the one dominating in simulations depending on subtle details of
the excitation processes.

But overall, it seems that the basic principals gleaned from the detailed
study of the standard, strong-confinement case can be quite logically
generalized to understand the trends in properties seen from this mosaic of
models spanning a broad range of rotation and
confinement parameters.

\section{SUMMARY \& FUTURE WORK}

This paper examines the effects of field-aligned rotation on
the magnetic channeling and confinement of a radiatively driven
stellar wind.
It builds upon the non-rotating models of Paper I, extending them to
cases of much stronger magnetic confinement (up to $\eta_{\ast} =
1000$), and
comparing a full spectrum of models ranging from weak to strong
confinement at equatorial rotation rates from zero to a substantial
fraction
($W=$~1/4 and 1/2) of the critical (or orbital) limit.
As an initial study, it ignores the effects of oblateness, gravity darkening
and limb darkening, and is based on an idealization of
isothermal flow driven by a purely radial line-force.

The key results can be summarized as follows:
\begin{itemize}

\item
The 2-D parameter space represented by rotation $W$ and magnetic
confinement $\eta_{\ast}$ can be conveniently divided by
comparing the Kepler radius $R_{\rm{K}} \approx W^{-2/3} \Rstar$
with the Alfv\'{e}n radius $R_{\rm{A}} \approx \eta^{1/4} \Rstar$.

\item
Models with $R_{\rm{A}} < R_{\rm{K}}$ have weak rotation and/or magnetic
confinement, with the effects of rotation limited to some
modest enhancement in equatorial density and overall mass loss rate,
relative to non-rotating cases.
In general any magnetically confined material falls back to the star.

\item
Transition models with $R_{\rm{A}} \approx R_{\rm{K}}$ have a complex
combination of inner region infall and outer region breakout,
but show no signs of accumulation of material into the kind of
extended Keplerian disk posited in the magnetically torqued disk
paradigm of Cassinelli et al. (2002).

\item
In models with $R_{\rm{A}} > R_{\rm{K}}$ the strong magnetic confinement
combines with a sufficiently rapid rotation that can support material
against infall near and above the Kepler radius, with then the
magnetic field both holding material in nearly rigid-body rotation,
and keeping it confined against the tendency for the
centrifugal force to propel material outward against gravity.

\item
Such strongly confined rotation models show a clear accumulation into
a {\em rigid-body} disk, much as predicted in the analytic Rigidly
Rotating Magnetosphere (RRM) formalism of TO-05.

\item
However, the present MHD simulations show that such rigid-body disks can be
highly dynamic and variable, with mass accumulation regulated by a
complex combination of inner disk infall and outer disk breakout.

\item
Nonetheless, application of the simple breakout analysis
introduced by TO-05 -- slightly modified to identify the outer disk
limit with the maximum radius of loop closure -- provides a quite
good, semi-quantitive agreement with both the breakout timescale and
limiting disk mass of the MHD simulations.

\item
The breakout events here are similar to those discussed by
ud-Doula, Townsend, and Owocki (2006), except the assumption here of
isothermal flow precludes us from following that paper's specific application
to modelling X-ray flares.
\end{itemize}

There thus remains much work for future extensions of the present
study, including relaxation of the assumptions of isothermal flow,
purely radial driving, 2-D axisymmetry with aligned dipole field, and
moderate rotation rates for which stellar oblateness can be neglected.

To more fully test the RRM paradigm, there is also a need to extend
MHD simulations even further into the large confinement limit,
to approach as closely as possible the estimated
$\eta_{\ast} \approx 10^{7}$ appropriate for Bp stars like
$\sigma$~Ori~E.

But even within the context the present set of models, we have
ignored here another key effect of magnetic rotation, namely the
outward angular momentum loss in the stellar wind, and
the associated spindown of the underlying star.
This omission was made to allow a more directed focus on the already
quite interesting, complex, and subtle effects of rotation on magnetic
confinement and disk formation.
But we have already carried out an extensive analysis of the implications
of our study for angular momentum loss and stellar spindown,
and so intend this to be the subject for the next paper in this series.

\section*{Acknowledgments}
This work was carried out with partial support by NASA Grants
Chandra/TM7-8002X and LTSA/NNG05GC36G, and by NSF grant AST-0507581.
We thank D. Cohen, M. Gagn\'{e}, D. Mullan, A.J. van Marle and A. Okazaki for many
helpful discussions.


\appendix

\section{Maximum Polar Alfv\'{e}n Speed}
\label{sec:appendix}

We derive here eqn.\ (\ref{vapmax}) cited in section  3.3,
giving a simple scaling relation between the expected maximum polar Alfv\'{e}n
speed and the  wind magnetic confinement parameter.
We begin by repeating eqn. (\ref{eta-eqn}) relating the ratio of
magentic to wind energy density to the ratio of Alfv\'{e}n speed to
flow speed,
\beq
\eta(r)\equiv \frac {B^2/8\pi}{\rho v^2/2}
= \left ( \frac{V_{\rm{A}}}{v} \right )^{2}
\, .
\label{eta-eqna}
\eeq
Here the density $\rho$ can be eliminated by noting that,
for a steady-state, magnetically channeled flow, the lack of
divergence in both the field and mass flux requires,
\beq
\frac{4 \pi \rho v}{ B}  = ~ {\rm constant}
\equiv \frac{{\dot M}}{B_{\rm{p}} \Rstar^{2}}
\, ,
\eeq
where we have normalized in terms of a polar surface field
$B_{\rm{p}}$ and global mass flux ${\dot M}$.
For a polar field that declines with radius as
$B = B_{p} (\Rstar/r)^{q}$,
this gives for the polar Alfv\'{e}n speed,
\beq
V_{\rm{Ap}}^{2} = \frac{B_{\rm{p}}\Rstar^{2}}{{\dot  M}} B v
= 4 \eta_{\ast}   \,
\left ( \frac{\Rstar}{r} \right )^{q}
v \vinf \, ,
\eeq
where the latter equality casts this in term of the magnetic
confinement parameter, with the factor 4 correcting for the fact that
our standard  $\eta_{\ast}$ is defined in terms of the equatorial surface
field $B_{eq}$, which is half the polar value $B_{p}$.
If we further assume a standard $\beta=1$ velocity law, then
\beq
V_{\rm{Ap}}
= 2 \sqrt{\eta_{\ast}}   \,  \vinf
\left ( 1 - \frac{\Rstar}{r} \right )^{1/2} \,
\left ( \frac{\Rstar}{r} \right )^{q/2}
\, .
\label{vapr}
\eeq
Setting the radial derivative of this to zero shows that the
maximum occurs at radius $r=(1+1/q) \Rstar$,
i.e. at $r=(4/3) \Rstar$ for the standard dipole case with $q=3$.
Plugging this radius into eqn.\ (\ref{vapr}), we find that the
expected maximum Alfv\'{e}n speed for dipole expansion of the polar
wind is
\beq
\max (V_{Ap} )
= \frac{\sqrt{27}}{8} \, \sqrt{\eta_{\ast}}   \,  \vinf
\approx 0.65 \, \sqrt{\eta_{\ast}}   \,  \vinf
\, .
\eeq
This thus provides a simple rule for the maximum Alfv\'{e}n speed to allow
in running MHD simulations with increasing magnetic confinement
parameter, as discussed in section 3.3.

\end{document}